\documentclass[
  reprint,       
  amsmath,amssymb,superscriptaddress,
  aps,
  prc
]{revtex4-2}

\usepackage{graphicx} 
\usepackage{dcolumn}  
\usepackage{bm}       
\usepackage{hyperref} 
\usepackage{braket}
\usepackage{comment}
\usepackage{float}

\begin{document}

\title{Improved quasiparticle nuclear Hamiltonians for quantum computing}

\author{Emanuele Costa}
\email{emanuele.costa@bsc.es}
\affiliation{Barcelona Supercomputing Center (BSC)}
\affiliation{Departament de Física Quàntica i Astrofísica, Universitat de Barcelona, 08028 Barcelona, Spain}
\affiliation{{Institut de Ciències del Cosmos, Universitat de Barcelona, 08028 Barcelona, Spain}}

\author{Javier Menéndez}
\email{menendez@fqa.ub.edu}
\affiliation{Departament de Física Quàntica i Astrofísica, Universitat de Barcelona, 
08028 Barcelona, Spain}
\affiliation{{Institut de Ciències del Cosmos, Universitat de Barcelona, 
08028 Barcelona, Spain}}

\date{\today}

\begin{abstract}
Quantum computing is increasingly offering concrete solutions toward the simulation of nuclear structure, with the potential to overcome the exponential scaling that limits classical diagonalization methods in large spaces. A particularly efficient encoding scheme, based on collective like-nucleon pairing modes, reduces the qubit requirements by half and avoids the non-local operator strings of standard fermion-to-qubit mappings. While this quasiparticle framework provides accurate results for semimagic nuclei, it does not adequately describe open-shell systems where proton-neutron correlations become important. In this work, we apply Brillouin-Wigner perturbation theory to systematically improve the quasiparticle description of open-shell nuclei in the $sd$ shell, reaching an energy relative error below $0.2\%$ compared to the nuclear shell model. Furthermore, to make the effective Hamiltonian suitable for quantum simulation, we introduce a mean-field Hartree-Fock approximation of the non-quasiparticle resolvent, achieving ground-state energies typically within $2\%$ of the exact shell-model result. This represents a systematic improvement over the bare quasiparticle Hamiltonian while remaining within the reach of near-term quantum devices. 
\end{abstract}

\maketitle

\section{Introduction}

Quantum computing is becoming a viable computational paradigm for addressing the complexity that limits classical numerical methods across various disciplines, from computer science and biology to chemistry and physics~\cite{RevModPhys.86.153,PRXQuantum.2.017003,Preskill2018quantumcomputingin,doi:10.1021/acs.chemrev.8b00803}. For instance, the central bottleneck of classical simulations of quantum many-body systems is the exponential growth of the Hilbert space with particle number, which renders exact solutions intractable beyond moderate system sizes. Quantum simulators and quantum computers circumvent this limitation by providing a native quantum framework for the representation and time evolution of many-body states, mapping the degrees of freedom of the physical system directly onto qubits and quantum gates~\cite{Daley2022,PhysRevA.99.052335,10.1098/rspa.1998.0162}. This natural mapping has been studied in a broad class of quantum systems that remain classically intractable, with promising results ranging from quantum chemistry \cite{shajan2024quantumcentricsimulationsextendedmolecules,Kandala2017,RevModPhys.92.015003} and condensed matter physics \cite{PhysRevB.102.235122,Yoshioka2024,PRXQuantum.5.037001} to high-energy \cite{PRXQuantum.4.027001,PhysRevResearch.2.023015} and nuclear physics~\cite{Zhang_2021,savage2023quantumcomputingnuclearphysics,stevenson2quantum,Ayral:2023ron}.

In the framework of low-energy nuclear physics, the impact of quantum computing is particularly relevant as the dimension of the nuclear many-body Hilbert space grows rapidly with nucleon number, rendering exact solutions prohibitively expensive for classical computers beyond the lightest systems. Variational quantum eigensolvers~\cite{Dumitrescu2018,Siwach:2021tym,PhysRevC.105.064317,PhysRevC.106.034325,Stetcu:2021cbj,perez2023nuclear,Sarma:2023aim,Bhoy_2024,PhysRevC.110.064320,zhang2025excited,Gu:2025yww,17n2-xh6k,Perez-Obiol:2024vjo,Aychet-Claisse:2025wos}, quantum projection techniques~\cite{PhysRevC.105.024324,PhysRevC.108.L031306,Rule:2024mrt}, imaginary time evolutions~\cite{Li:2023qdd} or quantum state preparation algorithms~\cite{costa2024quantum} have been proposed to study light and medium-mass nuclei, demonstrating the viability of quantum approaches for nuclear structure.
In addition, quantum algorithms have found applications in nuclear dynamics, reactions, and scattering~\cite{Weiss:2024mie,Singh:2025ubp,ZHANG2025139187,PhysRevC.109.064623,PhysRevC.111.034001,Rule:2026brk}. Furthermore, quantum computing tools have opened new avenues for nuclear theory through the lens of quantum information, enabling investigations of entanglement, quantum correlations, and nonstabilizerness in nuclear many-body systems \cite{PhysRevResearch.7.023228,Hengstenberg:2023ryt,Perez-Obiol:2023wdz,brokemeier2025quantum,Robin:2025wip,Robin:2025ymq}.

In the context of low-energy nuclear structure, the nuclear shell model (NSM)~\cite{RevModPhys.77.427,RevModPhys.92.015002,annurev:/content/journals/10.1146/annurev.ns.38.120188.000333,annurev:/content/journals/10.1146/annurev-nucl-101917-021120} is the preferred approach, but its quantum simulation remains challenging due to the complexity of nuclear interactions and the substantial overhead of fermion encodings. Several strategies have been proposed to reduce the latter~\cite{Gray-code-2,Stevenson25,Illa:2023scc,Robin:2023pgi,Du:2024zvr,Du:2025lhq}, 
including optimized qubit mappings, symmetry-adapted bases, and circuit compression techniques. Among these, quasiparticle encodings based on collective nucleon pairing modes~\cite{td9s-z7my,PhysRevC.109.064305,COSTA2026140042} have emerged as particularly promising: they exploit the dominance of pairing correlations in nuclei to reduce the number of qubits and replace non-local fermion-to-qubit operator strings with local qubit operators. This encoding has been successfully tested in variational quantum eigensolvers~\cite{td9s-z7my,PhysRevC.109.064305}, demonstrating significant reductions in both circuit depth and gate count. Furthermore, the local structure of quasiparticle Hamiltonians facilitates the implementation on analog quantum simulators, including Rydberg atom platforms, where the native interactions of the device can be directly matched to the quasiparticle operator algebra~\cite{chen2024spectroscopyelementaryexcitationsquench}.
However, in general the quasiparticle approximation achieves computational efficiency at the cost of reduced accuracy. It performs well for semimagic nuclei and certain neutron-rich isotopes, where pairing correlations dominate and the quasiparticle subspace captures the essential physics of the ground state. In these cases, the associated error remains comparable to typical algorithmic errors on current noisy intermediate-scale quantum (NISQ) devices \cite{bharti2022noisy,preskill2018quantum}. In contrast, for open-shell nuclei with similar numbers of protons ($Z$)  and neutrons ($N$), proton-neutron correlations beyond the pairing channel contribute significantly, and the quasiparticle approximation can introduce energy errors exceeding $10\%$~\cite{COSTA2026140042}. This severely limits the applicability of the quasiparticle framework across the nuclear chart. How to extend the accuracy of this approach  without sacrificing its computational advantages is therefore a key open question in nuclear quantum simulation.

In this work, we propose the application of Brillouin-Wigner (BW) perturbation theory~\cite{Brillouin1932LesPD,PhysRev.94.77,brillouin1,brillouin3,Shavitt_Bartlett_2009,brillouin2} — a method well established in nuclear physics, condensed matter, and particle physics — to systematically extend the accuracy of the quasiparticle approximation to open-shell nuclei. Within this framework, states outside the quasiparticle subspace are treated as virtual excitations, whose effect is folded into an energy-dependent effective Hamiltonian acting solely within the quasiparticle sector. This approach recovers ground-state energies with controlled and systematically improvable accuracy. To construct effective Hamiltonians suitable for near-term quantum devices, we introduce a mean-field approximation that replaces the correlated non-quasiparticle-sector propagator with a Hartree-Fock (HF) reference state~\cite{KELLY19676,PhysRev.46.618}. 
This approximation reduces the classical preprocessing cost 
while preserving the local qubit structure of the quasiparticle Hamiltonian, making it suitable for near-term quantum hardware.

The remainder of this article is organized as follows. Section~\ref{sec:nsm hamiltonian and quasiparticle}, introduces the NSM Hamiltonian and the quasiparticle pairing encoding. In Sec.~\ref{sec:brillouin wigner}, we derive the BW effective Hamiltonian within the quasiparticle sector, describe the self-consistent iterative algorithm, and analyze its computational cost. In Sec.~\ref{sec:bwi}, we introduce 
the truncated HF-BW method and discuss its physical motivation, operator structure, and scaling properties. In Sec.~\ref{sec:results}, we benchmark the accuracy of the BW and HF-BW methods against exact shell-model diagonalization for  $sd$-shell nuclei, assessing convergence, ground-state energy errors, state fidelity, and operator distance. Section~\ref{sec:conclusions}, summarizes the main findings and discusses perspectives for extensions to larger valence spaces and quantum hardware implementations.

\section{Quasiparticle pair Nuclear shell model Hamiltonian}
\label{sec:nsm hamiltonian and quasiparticle}

The NSM captures the dynamics of protons and neutrons within the nuclear valence space through the Hamiltonian:
\begin{equation}
    H_\text{CI}=\sum_a e_a c^{\dagger}_a c_a+ \frac{1}{4}\sum_{abcd} v_{abcd}c^{\dagger}_a c^{\dagger}_b c_d c_c,
    \label{eq:NSM}
\end{equation}
where $e_a$ are the single-particle energies 
and $v_{abcd}$ the two-body matrix elements encoding nucleon-nucleon interactions. The latter can be derived from chiral Hamiltonians~\cite{annurev:/content/journals/10.1146/annurev-nucl-101917-021120,PhysRevC.91.064301,PhysRevLett.113.142502}, or supplemented with phenomenological adjustments to better reproduce experimental data~\cite{RevModPhys.77.427,RevModPhys.92.015002}. The label $\text{CI}$ stands for active-space configuration interaction, referring to the many-body problem restricted to the valence space spanned by the active single-particle orbitals, in direct analogy with the active-space methods widely used in quantum chemistry~\cite{doi:10.1021/acs.jctc.3c00123}.  The operators $c^{\dagger}_a$ and $c_a$ are the fermion creation and annihilation operators for the nucleon mode $a$, respectively. Each mode $a$ labels a single-particle state in the valence space, $\ket{a}=\ket{n_a,l_a,j_a,m_a,t_a,t_{z,a}}$, where $n_a$ is the principal quantum number, $l_a$ the orbital angular momentum, $j_a$ the total angular momentum, $m_a$ its projection onto the quantization axis, and $t_{a}=1/2$ and $t_{z,a}$ are the isospin and its third component, distinguishing protons from neutrons.

A natural many-body basis for the NSM is given by the set of Slater determinants $\ket{s}$, built as antisymmetrized products of single-particle states according to the Pauli exclusion principle. These states take the explicit form 
\begin{align}
 \ket{s}=\bigotimes_{a \in I_{N}} c^{\dagger}_a \bigotimes_{b \in I_{Z}} c^{\dagger}_b \ket{\Omega}\,,
\end{align}
where $\ket{\Omega}$ denotes the core vacuum — the inert fully occupied core below the valence space — and $I_N$, $I_Z$ are the index sets of occupied single-particle states for valence neutrons and protons, respectively. By construction, each Slater determinant is an eigenstate of the total magnetic quantum number $M = \sum_a m_a$, a conserved quantity of $H_\text{CI}$. Without loss of generality in nuclei with even $N+Z$, we restrict to the $M = 0$ subspace.

For a nucleus with $N_n$ valence neutrons and $Z_p$ valence protons, the many-body Hilbert space has dimension $\dim(\mathcal{H}) = \binom{D}{Z_p} \times \binom{D}{N_n}$. Here, $D = \sum_{j \in J}(2j+1)$ is the number of available single-particle states per species, with $J$ the set of orbitals — groups of states sharing the same $n$, $l$, and $j$ but differing in $m_a$.
In this work, we focus on the \emph{sd} shell ($D=12$), which comprises the proton and neutron $0d_{5/2}$, $1s_{1/2}$, and $0d_{3/2}$ orbitals, and use the USDB Hamiltonian~\cite{Brown2006} as $H_\text{CI}$. The combinatorial growth of the Hilbert space with nucleon number makes the NSM computationally challenging for heavier nuclei, motivating the use of quantum computing approaches. 

However, the complexity of $H_\text{CI}$ and the fermion nature of the NSM degrees of freedom  make encoding onto digital or analog quantum devices highly demanding. A strategy to reduce this overhead is to project the fermion many-body Hilbert space onto a subspace spanned by collective fermion pairs, as discussed in Refs.~\cite{COSTA2026140042,td9s-z7my,PhysRevC.109.064305}. To this end, one introduces the quasiparticle pair operator
\begin{equation}
    Q^{\dagger}_A = c^{\dagger}_{a} c^{\dagger}_{\tilde{a}},
\end{equation}
where $\tilde{a}$ denotes the time-reversed partner of mode $a$, sharing all quantum numbers except for the magnetic projection, $m_{\tilde{a}}=-m_a$. The composite label $A \equiv \lbrace j_a, t_{z,a}, m_a, -m_a \rbrace$ identifies the quasiparticle mode. Physically, $Q^{\dagger}_A$ creates a Cooper-like pair of nucleons coupled to $M=0$~\cite{del2025hybrid}, a structure  favored by nuclear pairing~\cite{RevModPhys.75.607,Rios2017_PairingSRC}. 
The quasiparticle Hamiltonian 
\begin{align}
    H_Q&=\frac{1}{2}\sum_{AB} g_{AB}^{(1)}\, S^{+}_A S^{-}_B  + \frac{1}{2}\sum_{AB} g_{AB}^{(2)}\, N_A N_B\,,
    \label{eq h_q qubit}
\end{align}
is given in terms of quasiparticle modes $A,B$, which correspond to pairs $(a,\tilde{a})$ and $(b,\tilde{b})$, with 
\begin{align}
g^{(1)}_{AB}&=2\left( e_{a}+e_{\tilde{a}}\right)\left(\delta_{a,b}\delta_{\tilde{a},\tilde{b}}
-\delta_{\tilde{a},b} \delta_{a,\tilde{b}} \right)-v_{a,\tilde{a},b,\tilde{b}}\,, \\ 
g_{AB}^{(2)}&=v_{b,a,b,a}+v_{\tilde{a},b,\tilde{a},b}+v_{a,\tilde{b},a,\tilde{b}}+v_{\tilde{a},\tilde{b},\tilde{a},\tilde{b}}\,.
\end{align}

A first advantage is that the number of quasiparticle modes per orbital $j_a$ is $j_a + 1/2$, that is, half the degeneracy of the orbital. Concretely, the $0d_{5/2}$, $1s_{1/2}$, and $0d_{3/2}$ orbitals yield three, one, and two quasiparticle modes, respectively, thereby reducing the number of required qubits by half compared to a direct fermion encoding.

Second, pair operators satisfy commutation relations consistent with a hardcore boson algebra~\cite{doi:10.1142/S0217979206034947,10.21468/SciPostPhysLectNotes.82}:
\begin{align}
\left(Q^{\dagger}_A\right)^2 = 0, \, & & [Q_A, Q^{\dagger}_A] = 1 - 2N_A\,,
\end{align}
while operators acting on different modes commute. This algebraic structure admits a direct and local mapping onto qubit operators:
\begin{align}
    Q^{\dagger}_A = S^{+}_A, \, & & Q_A = S^{-}_A \,, & &N_A = Q^{\dagger}_A Q_A = \frac{1-Z_A}{2}\,,
\end{align}
where $S^{\pm} = \frac{1}{2}(X \pm iY)$ and $X, Y, Z$ are the single-qubit Pauli matrices. A second key advantage of this mapping is that it avoids the non-local Pauli strings that arise in standard fermion-to-qubit transformations such as Jordan-Wigner~\cite{jw} or Bravyi-Kitaev~\cite{bk} encodings, where operators acting on a single fermion mode can involve strings of Pauli operators that span the entire register. Here, each quasiparticle mode maps onto a single qubit with strictly local operators, a highly favorable feature for near-term quantum hardware.


Reference~\cite{COSTA2026140042} has analyzed the accuracy of the quasiparticle representation. $H_Q$ faithfully reproduces the $H_{\mathrm{CI}}$ ground state for semimagic nuclei ($N_n = 0$ or  $Z_p= 0$), where pairing correlations dominate and the quasiparticle subspace captures the essential physics. For open-shell nuclei ($N_n, Z_p >0$), the accuracy degrades in both ground state energy and fidelity due to significant proton-neutron correlations beyond pairing, especially in deformed systems.

This behavior motivates a systematic improvement of the quasiparticle method. Corrections arising from virtual excitations out of the quasiparticle subspace can be incorporated via an effective Hamiltonian built within BW perturbation theory~\cite{Brillouin1932LesPD,PhysRev.94.77,brillouin1,brillouin2,Shavitt_Bartlett_2009,brillouin3}. In this approach, the Hilbert space is partitioned into the quasiparticle model space and its orthogonal complement, with the effect of excluded states folded into renormalized matrix elements within the model space. The effective quasiparticle Hamiltonian retains the compact qubit representation while systematically accounting for many-body correlations beyond the bare pairing approximation.

\begin{figure*}[t]
    \centering
    \includegraphics[width=\textwidth]{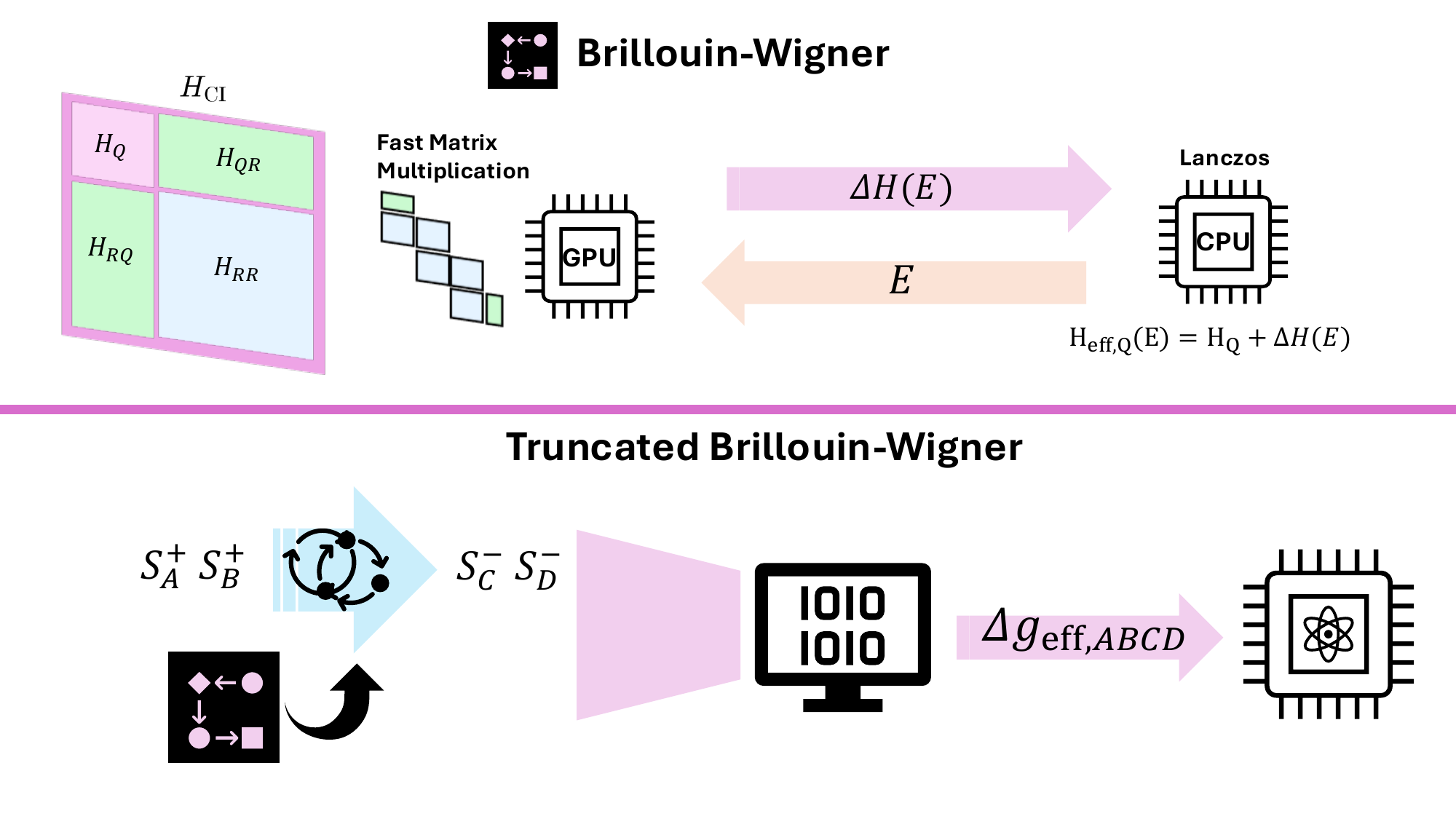}
    \caption{Schematic representation of the computational workflows for quasiparticle BW methods. \textit{Top panel:} In the full BW method, $H_\mathrm{CI}$ is decomposed into quasiparticle ($H_Q$), complementary ($H_{RR}$), and coupling ($H_{QR}$, $H_{RQ}$) blocks. The iterative BW algorithm evaluates the effective correction $\Delta H(E)$ via fast sparse matrix multiplications offloaded to GPU accelerators, while the Lanczos diagonalization of $H_{\mathrm{eff}}(E) = H_Q + \Delta H(E)$ determines the energy $E$ with CPUs. The self-consistent loop exchanges the updated $E$ and $\Delta H(E)$ until convergence. \textit{Bottom panel:} In the truncated BW method, the self-consistency loop is solved classically to obtain the converged effective two-body coupling $\Delta g_{\mathrm{eff},ABCD}$, which is then used to assemble the effective quasiparticle Hamiltonian expressed in terms of qubit operators $S^+_A S^+_B S^-_C S^-_D$. This Hamiltonian is passed to a quantum device for ground-state estimation, without any further classical-quantum feedback loop.}
    \label{fig0}
\end{figure*}

\section{Quasiparticle Brillouin-Wigner method}
\label{sec:brillouin wigner}

BW perturbation provides a systematic framework to derive an energy-dependent 
effective Hamiltonian acting solely within the quasiparticle model space $\mathcal{Q}$, spanned by the many-body basis
\begin{equation}
    \mathcal{B}_Q=\{ 
    \ket{s_Q}=\prod_{A \in I_Q} Q^{\dagger}_{A} \; \ket{\Omega} \},
\end{equation}
while virtual excitations account for the coupling to the complementary space $\mathcal{R}$. The fermion Fock space is partitioned as 
$\mathcal{F} = \mathcal{Q} \oplus \mathcal{R}$ via the orthogonal projectors
\begin{equation}
    \text{Q}=\sum_{s_Q \in \mathcal{B}_Q} \ket{s_Q}\bra{s_Q}, \qquad \text{R}=\text{I}-\text{Q}\,,
\end{equation}
where $\text{R}$ spans all Slater determinants 
with least two unpaired nucleons --- we work in the subspace $M=0$ --- such as modes with opposite $m$ but different $j$.

The top panel in Fig.~\ref{fig0} illustrates the computational workflow of the BW method. The Schrödinger equation 
\begin{align}
H_\text{CI}\ket{\Psi} = E\ket{\Psi}\,,    
\end{align}
is decomposed using $\ket{\Psi_Q} = \text{Q}\ket{\Psi}$ and $\ket{\Psi_R} = \text{R}\ket{\Psi}$, 
yielding the coupled system
\begin{align}
    H_{Q} \ket{\Psi_Q} + H_{\text{QR}} \ket{\Psi_R} &= E \ket{\Psi_Q}, \label{eq:Q_sector}\\
    H_{\text{RQ}} \ket{\Psi_Q} + H_{\text{RR}} \ket{\Psi_R} &= E \ket{\Psi_R}, \label{eq:R_sector}
\end{align}
where the projected Hamiltonians are
\begin{align}
H_{Q}=\text{Q} H_\text{CI} \text{Q}\,, & & H_{\text{QR}}=\text{Q} H_\text{CI} \text{R}\,, \nonumber \\ 
H_{\text{RQ}}=\text{R} H_\text{CI} \text{Q}\,, & & H_{\text{RR}}=\text{R} H_\text{CI} \text{R}\,. 
\end{align}
Formally solving Eq.~\eqref{eq:R_sector} for $\ket{\Psi_R}$ and substituting into 
Eq.~\eqref{eq:Q_sector} yields the effective Schrödinger equation
\begin{equation}
    H_{\text{eff}}(E)\ket{\Psi_Q} = E\ket{\Psi_Q}\,,
    \label{eq:eff_schrodinger}
\end{equation}
with the energy-dependent effective Hamiltonian
\begin{equation}
    H_{\text{eff}}(E) = H_{Q} + H_{\text{QR}}\left(E\,\text{I}_R - H_{\text{RR}}\right)^{-1} H_{\text{RQ}},
    \label{eq:Heff_exact}
\end{equation}
where $\text{I}_R$ is the identity restricted to $\mathcal{R}$. Unlike Rayleigh-Schrödinger 
perturbation theory \cite{Besalu1994,Shavitt_Bartlett_2009} ---- widely used in nuclear theory~\cite{Hjorth-Jensen:1995zrg,Coraggio:2008in} ---, the BW resolvent is evaluated at the exact eigenvalue $E$, rendering 
Eq.~\eqref{eq:eff_schrodinger} nonlinear and necessitating a self-consistent solution. 
We expand the resolvent as a geometric series,
\begin{equation}
    H_{\text{eff}}(E) = H_{Q} + \frac{1}{E}\sum_{n=0}^{N_\text{it}-1} H_{\text{QR}}
    \left(\frac{H_{\text{RR}}}{E}\right)^{\!n} H_{\text{RQ}} + \mathcal{O}\left(\frac{H_{\text{RR}}^{N_\text{it}}}{E^{N_\text{it}}}\right),
    \label{eq:Heff_series}
\end{equation}
which converges if the spectral radius satisfies $\rho(H_{RR}) < |E|$ \cite{dunford1988linear}. In practice, 
this condition is verified a posteriori for each nucleus, see Sec.~\ref{sec:results}. If the condition is satisfied, the ratio 
$\left(\rho(H_{RR})/|E|\right)^{N_\text{it}}$ controls the truncation error, which decreases geometrically with $N_\text{it}$.

\subsection{Computational implementation}

Since $H_{\text{eff}}(E)$ depends implicitly on the energy $E$, we solve Eq.~\eqref{eq:eff_schrodinger} iteratively. Starting from an initial estimate $E^{(0)}$ — taken here as the ground-state energy of $H_{Q}$ — at each iteration $k$ we obtain $H_{\text{eff}}(E^{(k)})$ by adding the matrix products in Eq.~\eqref{eq:Heff_series} up to order $N_\text{it}$, diagonalize the resulting matrix in $\mathcal{Q}$, and update the energy with its lowest eigenvalue $E^{(k+1)}$. We declare convergence when
\begin{equation}
    \left|E^{(k+1)} - E^{(k)}\right| < \delta\,,
\end{equation}
where $\delta$ is a threshold value. Alternatively, we also terminate the iteration if $E^{(k+1)} > E^{(k)}$, which signals an instability of the convergence.
Upon convergence, the lowest eigenvalue $E^{(k)}$ approximates the exact ground-state energy $E_\text{gs}$ of $H_\text{CI}$, with residual errors arising from the finite truncation order $N_\text{it}$.

From the standpoint of operator structure, each order $n$ in Eq.~\eqref{eq:Heff_series} generates induced multi-body quasiparticle interactions within $\mathcal{Q}$. This is, although $H_\text{CI}$ contains at most two-body interactions, successive insertions of the resolvent produce effective three-, four-, and higher-body terms, ultimately reaching up to $(N_n+Z_p)/2$-body interactions. 
This feature 
represents a fundamental obstacle to a direct implementation on quantum devices, where simulating high-body operators significantly increases circuit depth and gate count.

Nonetheless, clasically, the quasiparticle BW method can be advantageous, because rather than diagonalizing $H_\text{CI}$ in the full space $\mathcal{F}$ of dimension $\dim\mathcal{F}$, one performs $N_\text{it}$ diagonalizations of $H_{\text{eff}}$ in $\mathcal{Q}$, with $\dim\mathcal{Q} \ll \dim\mathcal{F}$. 
The dominant overhead is the evaluation of the matrix products 
$H_{\text{QR}}H_{\text{RR}}^{n}H_{\text{RQ}}$ at each order, which scales as $\mathcal{O}(N_\text{it} \cdot \dim\mathcal{Q} \cdot \dim\mathcal{F}^2)$. These products can be efficiently computed on GPU accelerators \cite{10.1145/1654059.1654078,Dongarra2014}.

\section{Truncated quasiparticle Brillouin-Wigner method with Hartree-Fock ansatz}
\label{sec:bwi}

To get an efficient encoding of effective quasiparticle Hamiltonian suitable for near term quantum devices, we truncate the BW method to the one-quasiparticle-proton--one-quasiparticle-neutron subspace, the quasiparticle subspace of the $N_n=Z_p=2$ Fock sector. In the \emph{sd} shell, this sector describes $^{20}\mathrm{Ne}$. In this way we get an effective interaction
\begin{align}
    H_{\mathrm{eff,tr}} = H_{Q} + \Delta H^{(N,P)}_{\mathrm{eff}}\,,
\label{eq:delta_H_trun}    
\end{align}
where
\begin{align}
     \Delta H^{(N,P)}_{\mathrm{eff}}&=\frac{1}{4}\sum_{ABCD} \Delta g_{\mathrm{eff},ABCD}\, S^{+}_AS^{+}_B S^{-}_C S^{-}_D\,, \label{eq:delta_H} \\
    \Delta g_{\mathrm{eff},ABCD}&=\bra{AB}\frac{1}{E^*} H^{(2,2)}_{\text{QR}} \sum_{n=0}^{N_{\text{it}}} \frac{H^{(2,2),n}_{\text{RR}}}{E^{*,n}} H^{(2,2)}_{\text{RQ}} \ket{CD}\,,
\label{eq:delta_g}
\end{align}
with $E^*$ the energy obtained at the end of the self-consistent method. The superscript $(2,2)$ indicates that the projected blocks $H_{\text{QR}}$, $H_{\text{RR}}$, 
and $H_{\text{RQ}}$ are evaluated within the $N_n=Z_p=2$ sector. 

The bottom panel in Fig.~\ref{fig0} illustrates the computational workflow of the truncated BW method. The restriction to two-body quasiparticle interactions yields a Hamiltonian with a significantly reduced number of terms and local operator connectivity.
In particular, the absence of high-body interactions directly translates into shallower quantum circuits and reduced gate counts compared to a direct encoding of $H_\mathrm{CI}$, making $H_{\mathrm{eff,tr}}$ well-suited for near-term quantum devices.

While $H_\text {eff,tr}$ can be computed classically since $\dim \mathcal{F}^{(2,2)}=D^2(D-1)^2/4$, this still requires iterative sparse matrix multiplications that can become computationally expensive. To reduce this overhead further, we introduce two additional approximations: we only keep the ground-state pole in the resolvent, and evaluate it at the mean-field level.

We begin by expanding the spectral representation,
\begin{equation}
    (E-H^{(2,2)}_{RR})^{-1} = \frac{\ket{\Phi^{(2,2)}_0}\bra{\Phi^{(2,2)}_0}}{E-E^{(2,2)}_0}
     + 
    \sum_{i=1}^{M} \frac{\ket{\Phi^{(2,2)}_i} \bra{\Phi^{(2,2)}_i}}{E-E^{(2,2)}_i},
    \label{eq:spectral_decomp}
\end{equation}
where $\{\ket{\Phi^{(2,2)}_i}, E^{(2,2)}_i\}$ are the energy-ordered eigenstates and eigenvalues  of $H^{(2,2)}_{\text{RR}}$. 
Since $H_{Q}$ partially captures the low-energy spectrum, the pole associated with the ground state $\ket{\Phi^{(2,2)}_0}$ already provides a significant contribution to the BW correction. We therefore retain only the leading term in Eq.~\eqref{eq:spectral_decomp}, neglecting all excited-state contributions 
$i \geq 1$.

In addition, to avoid the cost of diagonalizing $H^{(2,2)}_{RR}$, we approximate its ground state 
with a Hartree-Fock calculation on $H_\mathrm{CI}$ 
restricted to $N_n=Z_p=2$, obtaining $\ket{\Psi^{(2,2)}_\mathrm{HF}}$, and 
project it onto $\mathcal{R}$ with normalization,
\begin{equation}
    \ket{\Psi^{(2,2)}_{\text{RR},\mathrm{HF}}} = 
    \frac{\text{R}\ket{\Psi^{(2,2)}_\mathrm{HF}}}
    {||\text{R}\ket{\Psi^{(2,2)}_\mathrm{HF}}||}\,,
    \label{eq:HF_projected}
\end{equation}
with associated energy pole
\begin{equation}
    E_\mathrm{HF}^{\text{RR},(2,2)} = \bra{\Psi^{(2,2)}_{\text{RR},\mathrm{HF}}} 
    H^{(2,2)}_{\text{RR}} \ket{\Psi^{(2,2)}_{\text{RR},\mathrm{HF}}}\,.
    \label{eq:E_HF_RR}
\end{equation}
By the variational principle, $E_\mathrm{HF}^{\text{RR},(2,2)}$ is an upper bound  to the true ground state energy $E^{(2,2)}_0$ of $H^{(2,2)}_{\text{RR}}$. 
Nevertheless, it is not the tightest possible mean-field upper bound — a tighter one could be obtained by a HF optimization within $\mathcal{R}$. The normalization in Eq.~\eqref{eq:HF_projected} ensures consistency with the rank-one projector structure of Eq.~\eqref{eq:spectral_decomp}, where $\ket{\Phi^{(2,2)}_0}$ is a normalized eigenstate. The density matrix associated with this reference state is
\begin{equation}
    \rho_{\text{RR},\mathrm{HF}}^{(2,2)} = \ket{\Psi^{(2,2)}_{\text{RR},\mathrm{HF}}}
    \bra{\Psi^{(2,2)}_{\text{RR},\mathrm{HF}}}\,.
\end{equation}

Approximating $\ket{\Phi^{(2,2)}_0} \approx \ket{\Psi^{(2,2)}_{\text{RR},\mathrm{HF}}}$, $E^{(2,2)}_0 \approx E_\mathrm{HF}^{\text{RR},(2,2)}$ in the BW expansion, the effective two-body quasiparticle matrix elements simplify to
\begin{equation}
    \Delta g^{\mathrm{HF}}_{ABCD} = \frac{\bra{AB} H^{(2,2)}_{QR}\, 
    \rho_{\text{RR},\mathrm{HF}}^{(2,2)}\, H^{(2,2)}_{RQ} \ket{CD}}
    {E^* - E_\mathrm{HF}^{\text{RR},(2,2)}},
    \label{eq:delta_g_HF}
\end{equation}
where $E^*$ is the converged energy obtained from the classical self-consistency loop.
Once convergence is reached, the matrix elements $\Delta g^\mathrm{HF}_{ABCD}$ are fully determined. Then, the effective Hamiltonian
\begin{equation}
    H_{\mathrm{eff,HF}} = H_{Q} + \frac{1}{4}\sum_{ABCD} 
    \Delta g^{\mathrm{HF}}_{ABCD}\, S^+_A S^+_B S^-_C S^-_D,
    \label{eq:Heff_HF}
\end{equation}
can be assembled once and passed directly to the quantum device for ground- state estimation. This is a significant practical advantage, as it places no limitation on quantum algorithms: any ground state solver, including variational quantum eigensolvers or quantum phase 
estimation \cite{mande2023tight,perez2023nuclear,td9s-z7my,PhysRevC.109.064305}, can be applied directly to $H_{\mathrm{eff,HF}}$.


Since $H_{\mathrm{eff,HF}}$ incorporates the leading mean-field contribution to the BW renormalization within the $(2,2)$ sector, it systematically captures more ground-state correlations compared to the bare quasiparticle Hamiltonian $H_{Q}$ 
with moderate additional quantum resources.

\subsection{Truncated quasiparticle Brillouin-Wigner beyond Hartree-Fock}
\label{sec:bw truncated}

For comparison, we can also solve the truncated BW quasiparticle method based on Eqs.~\eqref{eq:delta_H_trun}, \eqref{eq:delta_H} and~\eqref{eq:delta_g} without the HF approximation. 

This increases the classical computational cost of the truncated BW approach. At the HF level, 
beyond the initial HF calculation in the $(2,2)$ sector, the dominant cost is the single evaluation of $H^{(2,2)}_{\text{QR}}\ket{\Psi^{(2,2)}_{\text{RR},\mathrm{HF}}}$, of dimension $d \times D_Q^2(D_Q-1)^2/4$, where $d$ is the number of nonzero elements per row of $H^{(2,2)}_{\text{QR}}$ and $D_Q$ the number of quasiparticle modes. The outer product structure of Eq.~\eqref{eq:delta_g_HF} allows $\Delta g^\mathrm{HF}_{ABCD}$ 
to be assembled with a single matrix-vector multiplication, making the total 
classical preprocessing cost relatively low. 

Going beyond the HF level introduces an additional challenge. By construction, the matrix elements in Eq.~\eqref{eq:delta_g} encode virtual 
excitations out of the two-quasiparticle sector only, and therefore do not account for the Pauli principle due to the remaining $N_n + Z_p - 4$ nucleons. Neglecting these constraints overestimates  the energy corrections, since transitions into already-occupied single-particle states are incorrectly permitted. To mitigate this, we introduce a statistical Pauli blocking factor $P(s_F, s_I)$ into the projected Hamiltonian blocks,
\begin{align}
    H^{(2,2)}_{\text{RQ}}=\sum_{RQ} \bra{s_R} H_{\text{RQ}} \ket{s_Q} P(s_R,s_Q) \ket{s_R} \bra{s_Q}\,, \nonumber \\
    H^{(2,2)}_{\text{RR}}=\sum_{RR'} \bra{s_R} H_{\text{RR}} \ket{s_{R'}} P(s_R,s_{R'}) \ket{s_{R}} \bra{s_{R'}}\,,
    \label{eq:pauli_blocking}
\end{align}
with $H^{(2,2)}_{\text{QR}} = \left(H^{(2,2)}_{\text{RQ}}\right)^\dagger$. We define the blocking factor as the combinatorial probability that all single-particle states involved in a transition $s_I \to s_F$ are unoccupied,
\begin{equation}
    P(s_{F},s_{I})=\prod_{i \in  s_{F} / s_{I}}\left(1 -\frac{(N_ i-1)}{D}\right)\,.
\end{equation}
Here, the product runs over the single-particle states that are newly occupied in $s_F$ relative to $s_I$, and $N_i \in \{N_n, Z_p\}$ is the total number of valence nucleons corresponding to mode $i$. Each factor $(1-(N_i-1)/D)$ estimates the probability that a given state is not occupied by any of the remaining $N_i - 1$ nucleons of the same species, in a mean-field approximation within the shell.

%


\section{Results}
\label{sec:results}

\subsection{Quasiparticle Brillouin-Wigner in the $sd$ shell}

\begin{figure}[t]
    \centering
    \includegraphics[width=1\linewidth]{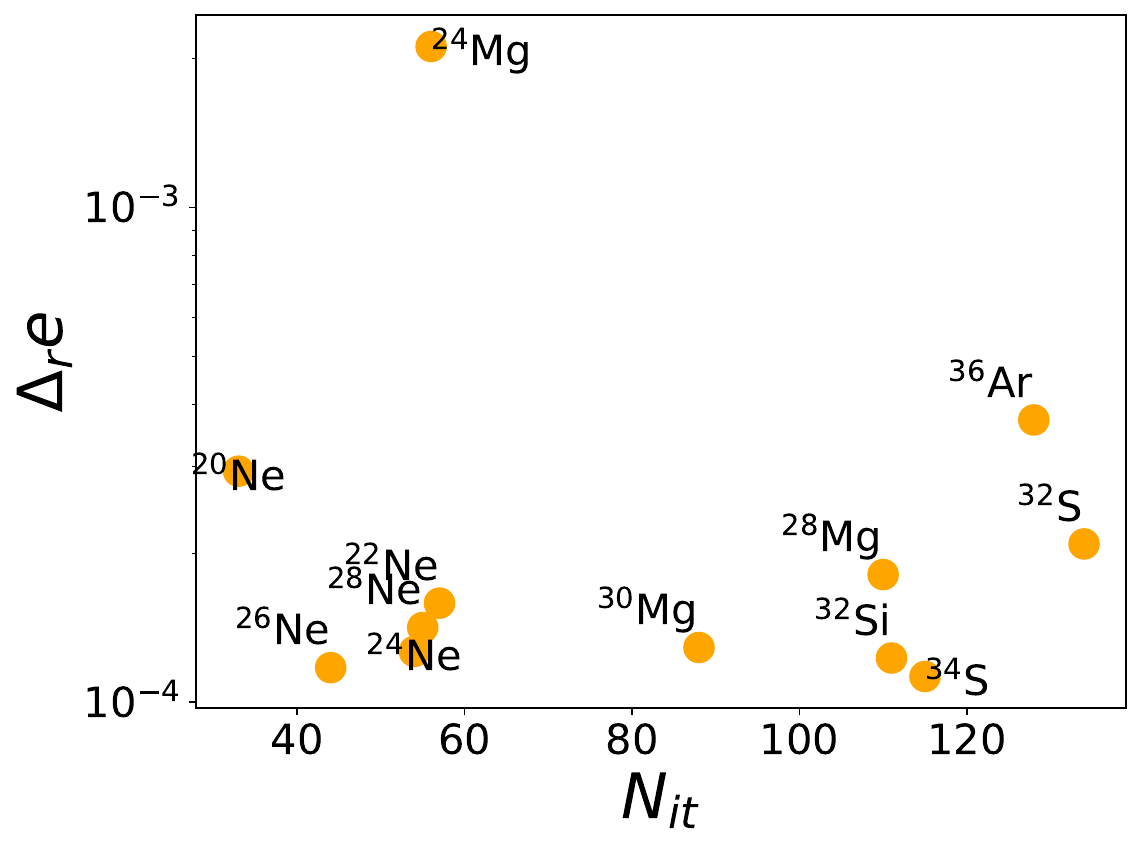}
    \caption{Relative energy error $\Delta_r e = (E^* - E_\mathrm{gs})/|E_\mathrm{gs}|$ 
    at convergence of the full BW method as a function of the total number of iterations 
    $N_\mathrm{it}$, for \emph{sd}-shell nuclei.}
    \label{fig:1}
\end{figure}

First, we study the convergence of the full quasiparticle BW method for a representative subset of even-$N$--even-$Z$ $sd$-shell nuclei, ranging from $^{20}$Ne to $^{36}$Ar. We fix the convergence threshold to $\delta = 10^{-3}$.

Figure~\ref{fig:1} shows the residual relative error compared to the exact shell-model result, $\Delta_r e = (E^* - E_\mathrm{gs})/|E_\mathrm{gs}|$, at convergence as a function of the total number of iterations needed to reach convergence. Figure~\ref{fig:1} highlights that for all nuclei considered, the full BW method converges to a relative error at the $\Delta_r e \sim 10^{-4}$ level.  The only larger error, $\Delta_r e \sim 2\cdot10^{-3}$, appears for $^{24}$Mg.
This nucleus is characterized by a strong contribution from non-quasiparticle configurations and significant deformation. This complexity has been previously observed using quantum algorithms~\cite{costa2024quantum} and quantum information analyses~\cite{brokemeier2025quantum}. These results demonstrate the potential of the quasiparticle BW method to capture additional nuclear correlations beyond pairing.

\begin{figure}[t]
    \centering
    \includegraphics[width=1\linewidth]{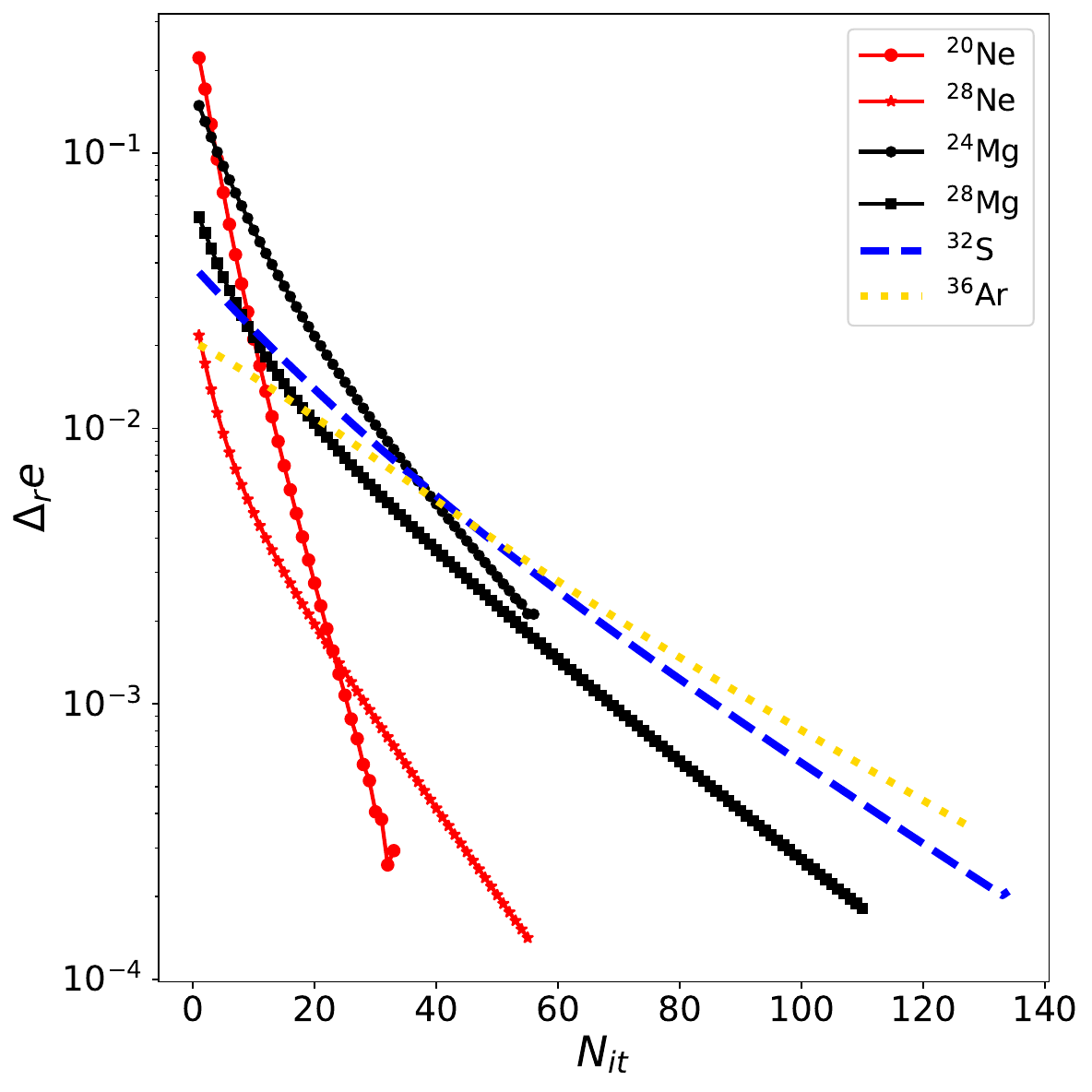}
    \caption{Relative energy error $\Delta_r e$ convergence as a function of iteration number $N_\mathrm{it}$, for $N=Z$ $^{20}$Ne (solid red line with circles), $^{24}$Mg (solid black line with circles), $^{32}$S (blue dashed line), $^{36}$Ar (yellow dotted line), and neutron-rich $^{28}$Ne (solid red line with triangles), $^{28}$Mg (solid black line with squares). 
    }
    \label{fig:2}
\end{figure}

Comparing Ne isotopes, neutron-rich $^{26}$Ne and $^{28}$Ne, despite having more valence nucleons, require a similar number of iterations and converge to a comparable relative energy error as lighter $^{22}$Ne and $^{24}$Ne. The isotopes of other elements --- excluding $^{24}$Mg --- show a similar behaviour in the compared number of iterations needed for convergence. Figure~\ref{fig:1} also shows that isotopes of heavier elements tend to converge after more iterations.

%
%

Figure~\ref{fig:2} illustrates in more detail the relative energy-error convergence of the full quasiparticle BW method for selected $sd$-shell nuclei, as a function of the number of iterations $N_\text{it}$. In all cases, the convergence of $|\Delta_r e|$ follows a stretched exponential pattern, $|\Delta_r e| \propto \exp(-\gamma N_\mathrm{it}^\alpha)$, where $\alpha$ and $\gamma$, which control the steepness of the convergence curve, depend on the nucleus. First, Fig.~\ref{fig:2} shows that deformed nuclei such as $^{20}$Ne and $^{24}$Mg, which are sensitive to nuclear correlations beyond pairing, are not well described by $H_Q$, the starting point of the curves in Fig.~\ref{fig:2}. In contrast, neutron-rich nuclei such as $^{28}$Ne start from a much better $|\Delta_r e|$ value at $N_\text{it}=1$. 

Regarding the convergence as a function of $N_\text{it}$, Fig.~\ref{fig:2} shows that, for $N_n = Z_p$ nuclei, the convergence curves become progressively less steep from lighter to heavier systems: starting from $^{20}$Ne and following $^{24}$Mg, $^{32}$S and $^{36}$Ar the convergence is slower, indicating that each BW iteration improves less the energy estimate for heavier nuclei. Likewise, within an isotope chain, the curves become flatter for neutron-rich nuclei, as illustrated by comparing $^{20}$Ne vs $^{28}$Ne or $^{24}$Mg vs $^{28}$Mg in Fig.~\ref{fig:2}. Both trends can be understood in terms of the perturbative coupling $\rho(H_{RR})/|E|$ that controls the BW series. With  more valence nucleons, the energy scale $|E|$ grows, reducing the effective coupling and weakening the contribution of virtual non-quasiparticle excitations to the BW corrections. This results in a slower convergence curve. 

Altogether, for the isotope chains in Fig.~\ref{fig:1}, neutron-rich nuclei are better described by $H_Q$, but they converge more slowly. Therefore, all isotopes along the  chain converge in a similar number of iterations. Heavier elements, which in the $sd$-shell do not permit very neutron-rich systems, converge more slowly than lighter ones.

\begin{figure}[t]
    \centering
    \includegraphics[width=1\linewidth]{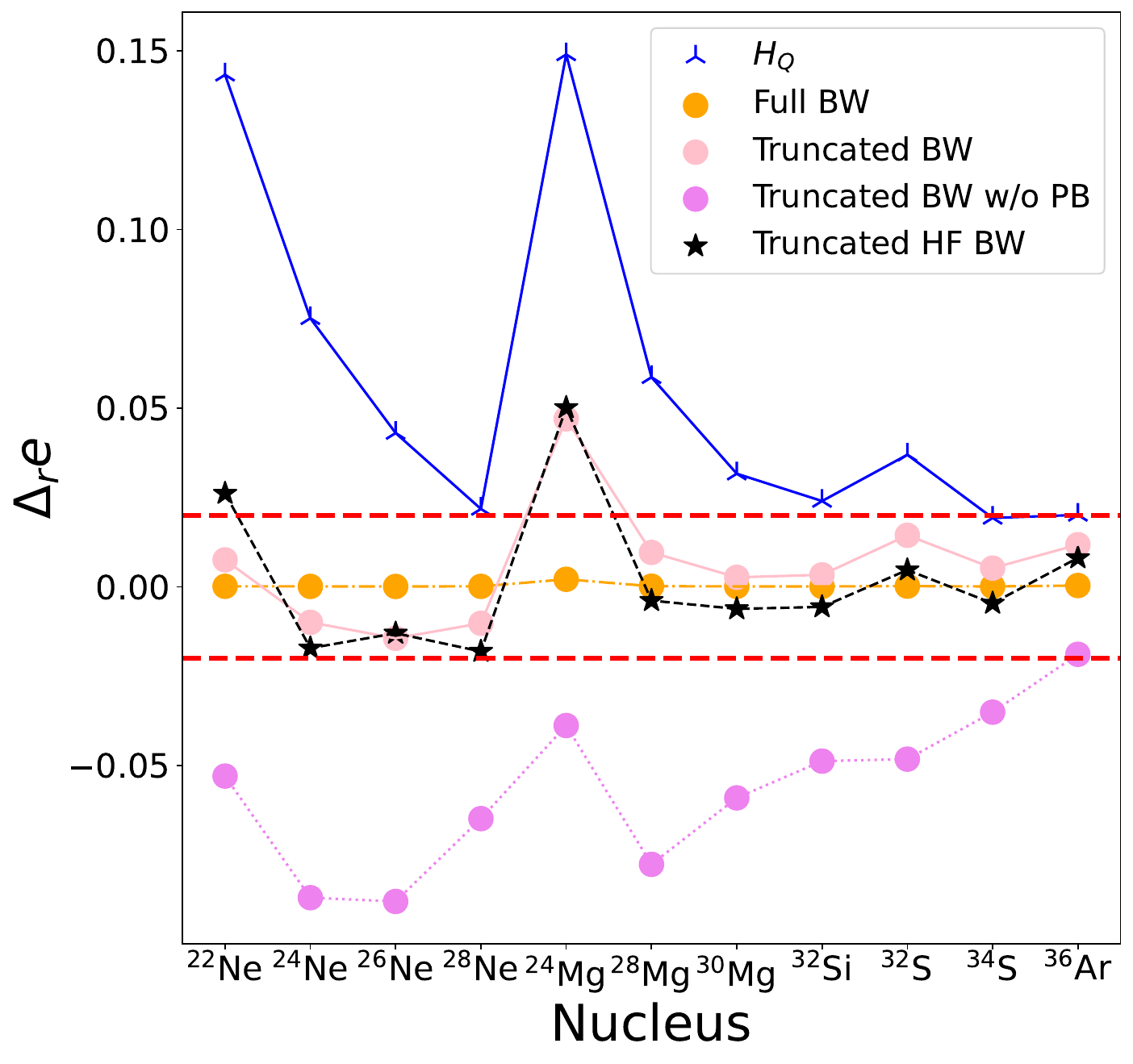}
    \caption{Relative energy error $\Delta_r e = (E^* - E_\mathrm{gs})/|E_\mathrm{gs}|$ for $sd$-shell nuclei, estimated by different methods: the quasiparticle Hamiltonian $H_Q$ (blue symbols), the full quasiparticle BW $H_{\text{eff}}$ (orange circles), the truncated HF-BW $H_{\text{eff},\text{HF}}$ (black stars), and the truncated BW $H_{\text{eff},\text{tr}}$ with (light magenta circles) and without (deep magenta circles) Pauli blocking. The red dashed lines indicate the $\pm 2\%$ accuracy threshold.}
    \label{fig:3}
\end{figure}

\subsection{Truncated Brilouin-Wigner: energy comparison}

Figure~\ref{fig:3} compares the ground-state energy of the quasiparticle BW method with the exact $H_\text{CI}$ shell-model exact result, via the relative error $\Delta_r e = (E^* - E_\mathrm{gs})/|E_\mathrm{gs}|$. The orange circles show the results for the full BW approach, corresponding to Fig.~\ref{fig:1}, which reproduces the exact energies accurately in all cases. In contrast, the bare $H_Q$ (blue symbols) systematically overestimates the ground-state energy, with errors ranging from $\Delta_r e \simeq 0.02$ for the most neutron-rich isotopes to $\Delta_r e \simeq 0.15$ for $^{22}$Ne and $^{24}$Mg~\cite{COSTA2026140042}. These differences highlight again the limitation of the bare quasiparticle approach.

The truncated BW method clearly improves on the $H_Q$ results. The black stars in Fig.~\ref{fig:3} show the relative energy errors for the truncated HF BW method. The results exclude $^{20}$Ne, which has $N_n=2,Z_p=2$ and therefore the truncated BW reduces exactly to the full BW method  --- giving a very good energy for this nucleus. The average error for the HF BW is just $\overline{\Delta_r e} \simeq 0.016$ and the majority of results lie within the $2\%$ threshold, indicated by the red dashed lines. This level of accuracy is similar to the the precision obtained in Ref.~\cite{COSTA2026140042} for semimagic nuclei using the bare quasiparticle Hamiltonian, $H_Q$. The exceptions are $^{22}$Ne with $\Delta_r e \simeq 0.026$ and $^{24}$Mg with $\Delta_r e \simeq 0.05$. $\Delta_r e$ appears to be rather constant within each isotope chain
Variational violations occur for most nuclei, but they remain within the $2\%$ threshold and we have checked that they are not caused by algorithmic instabilities.

The exceptions where the HF BW performs worse are $^{22}$Ne with $\Delta_r e \simeq 0.026$ and $^{24}$Mg with $\Delta_r e \simeq 0.05$. These nuclear structure of these two nuclei is dominated by correlations beyond pairing, as indicated by the poor results obtained with the quasiparticle $H_Q$. Unlike $^{20}$Ne, which is naturally well described with one-quasiparticle-nucleon--one-quasiparticle proton terms, efficiently capturing the additional correlations for these nuclei requires to extend the truncated BW method at the two-quasiparticle-nucleon ($^{22}$Ne) and two-quasiparticle-proton ($^{24}$Mg) level.   

Figure~\ref{fig:3} also indicates that the truncated BW method with Pauli blocking (light pink symbols) achieves a similar accuracy to the simpler HF BW, with $|\Delta_r e| \leq 2\%$ for all nuclei except $^{24}$Mg, where $\Delta_r e \simeq 0.047$.  The average error across the remaining nuclei is $\overline{\Delta_r e} \simeq 0.014$.
Here, the small violations of the variational principle can be attributed to the statistical Pauli blocking underestimating the suppression of the effective coupling for the corresponding nuclei. 
In contrast, the truncated BW without Pauli blocking (magenta circles) shows large and systematic variational violations, with $\Delta_r e$ ranging from $\simeq -0.019$ for 
$^{36}$Ar to $\simeq -0.086$ for $^{28}$Ne, well outside the 2\% level. This demonstrates that the Pauli blocking is essential for obtaining physically meaningful results in this case.

\subsection{Truncated Brilouin-Wigner: state and Hamiltonian fidelity}

\begin{figure}[t]
    \centering
    \includegraphics[width=1\linewidth]{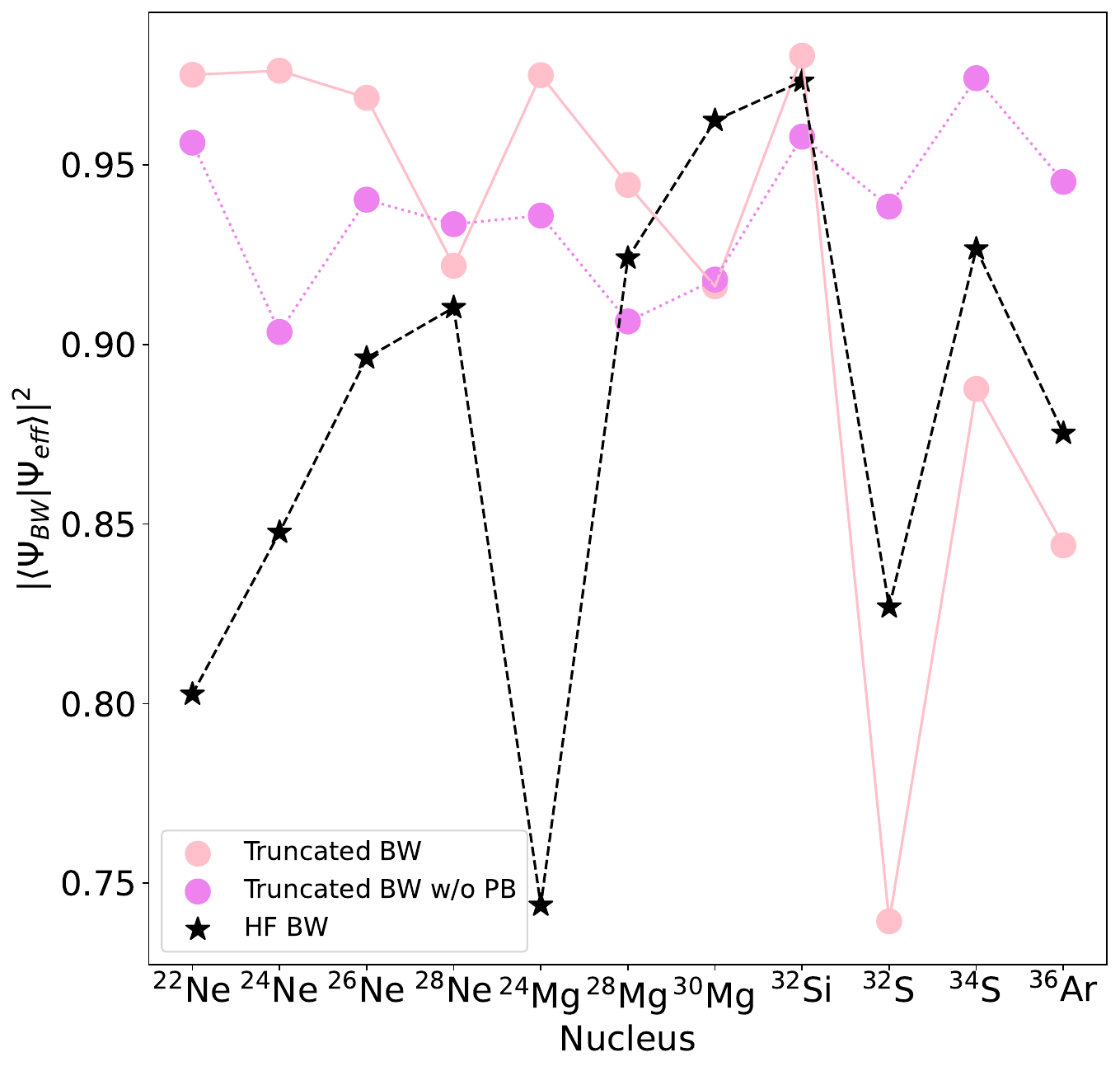}
    \caption{Fidelity $|\braket{\Psi_\mathrm{BW}|\Psi_\mathrm{eff}}|^2$ across $sd$-shell nuclei of the ground states of the truncated HF BW Hamiltonian, $H_{\mathrm{eff,HF}}$, and the truncated BW Hamiltonian, $H_{\mathrm{eff,tr}}$, with and without Pauil blocking, with respect to the ground state of the full BW effective Hamiltonian $H_\mathrm{eff}$. The symbol and color codes are the same as in Fig.~\ref{fig:3}.}
    \label{fig:4}
\end{figure}

To further analyze the quality of the ground states obtained with the truncated BW method, Fig.~\ref{fig:4} displays the fidelity $F=|\braket{\Psi_\mathrm{BW}|\Psi_\mathrm{eff}}|^2$ of these states with respect to ground state of the full BW effective Hamiltonian, $H_\mathrm{eff}$. We take the latter as a reference because it is the best approximation to the exact shell-model ground state of $H_\mathrm{CI}$ within the quasiparticle sector $\mathcal{Q}$.

Figure~\ref{fig:4} shows that the truncated HF BW method (black stars) has an average fidelity, $\overline{F}_\mathrm{HF} \simeq 0.88$, with sizeable difference between nuclei, $F_\mathrm{HF} \in [0.74, 0.97]$. Naturally, the two nuclei with larger $\Delta_r e$, $^{22}$Ne and $^{24}$Mg, show the lower fidelities, but also for $^{22}$S we find $F_\mathrm{HF}<0.85$. In an isotope chain, $F_\mathrm{HF}$ increases with neutron number, where proton-neutron correlations become less relevant. Overall, the HF WB fidelities in Fig.~\ref{fig:4} are comparable to those reported in Ref.~\cite{COSTA2026140042} for the ground state of $H_Q$ with respect to the exact CI ground state for semimagic nuclei, $F_Q \in [0.95, 1]$. Assuming that the ground state of $\ket{\Psi_\mathrm{BW}}$ provides a faithful representation of the $H_\mathrm{CI}$ ground state within $\mathcal{Q}$, this suggests that the truncated BW method recovers, for open-shell nuclei, a level of accuracy comparable to that achieved by the bare quasiparticle Hamiltonian for semimagic ones.

Figure~\ref{fig:4} also highlights that, in general, for Ne and Mg isotopes $F_\mathrm{HF} < F_\mathrm{tr}$, the fidelity of the truncated BW method beyond HF including Pauli blocking (light pink circles), while for the S isotopes and $^{32}$Si the HF BW method provides higher fidelities. Overall, the truncated BW with Pauli blocking achieves fidelities $F_\mathrm{tr} \in [0.74, 0.98]$ with average $\overline{F}_\mathrm{tr} \simeq 0.92$. However, 
$^{24}$Mg displays a high fidelity despite its large energy discrepancy, and
the fidelity is very low for $^{32}$S and $^{36}$Ar, where the energy accuracy is high. Likewise, the truncated BW without Pauli blocking (magenta circles in Fig.~\ref{fig:4}) achieves very high fidelities similar or even better than the ones of the other truncated BW calculations,
which is in contrast with the poor energies predicted for all these nuclei. 
These inconsistencies between fidelity and energy error for $^{24}$Mg and the truncated BW without Pauli blocking suggest that the energy estimation is sensitive to the operator structure of the effective Hamiltonian rather than to the overlap of the ground states alone. 

\begin{figure}[t!]
    \centering
    \includegraphics[width=1\linewidth]{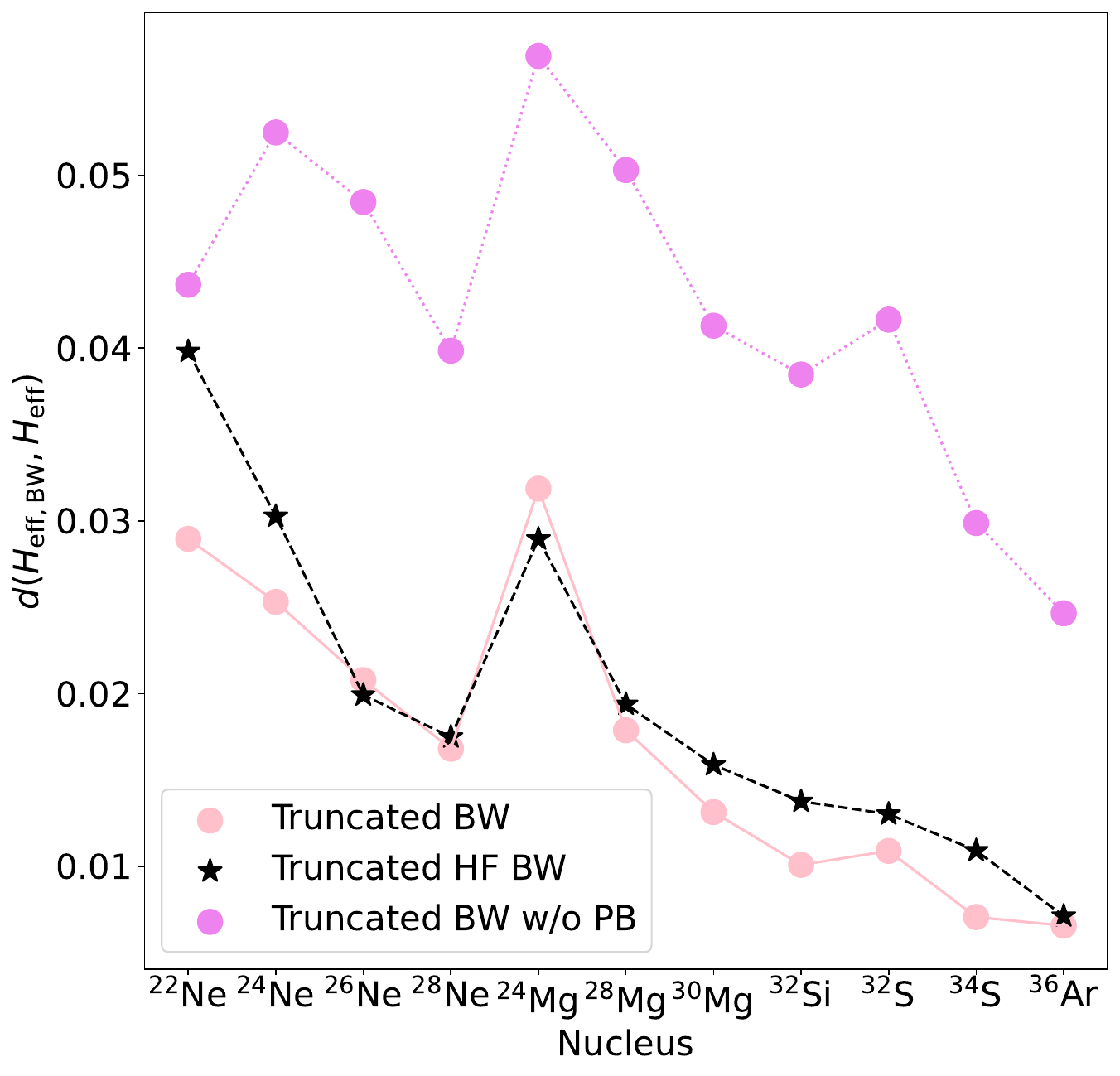}
    \caption{Normalized operator distance $d(H_{\mathrm{eff,BW}}, H_\mathrm{eff}) = ||H_{\mathrm{eff,BW}} - H_\mathrm{eff}||/||H_\mathrm{eff}||$ with respect to the full BW effective Hamiltonian, for the truncated HF BW effective Hamiltonian, and for the truncated BW one with and without Pauli blocking. Results for the same nuclei and with the same symbol and color codes as in Figs.~\ref{fig:3} and~\ref{fig:4}.}
    \label{fig:5}
\end{figure}

To quantify this, Fig.~\ref{fig:5} shows the normalized $L_2$ operator distance $d(A,B) = ||A - B||/||A||$ between the truncated effective Hamiltonian and the full BW one. The truncated HF-BW method (black stars) show operator distances in the range $[0.007, 0.032]$, which are smaller for heavier nuclei. These BW HF distances are similar than for the truncated BW with Pauli blocking (light pink circles), but much smaller compared to the BW without Pauli blocking, which has $[0.025, 0.057]$, consistently with its poor energies. Overall, the results in Fig.~\ref{fig:5} establish that the operator distance $d(H_{\mathrm{eff,tr}}, H_\mathrm{eff})$ is a more reliable diagnostic of method accuracy than the state fidelity alone. Finally, Fig.~\ref{fig:5} also shows that the operator distance is anomalously large for $^{24}$Mg. This suggests that its complexity within this framework is associated with off-diagonal operator contributions that do not scale with nucleon number in the same way as the diagonal terms. 

\section{Conclusion}
\label{sec:conclusions}

In this article, we introduce the BW method as a way to improve the reach of quasiparticle pair Hamiltonians beyond semimagic nuclei within the nuclear shell model. In particular, we present the quasiparticle BW method as a classical algorithm
and simplify it by adopting a two-body quasiparticle truncation and a HF ansatz for the non-quasiparticle propagator, obtaining effective Hamiltonians with polynomial preprocessing cost. The locality of the resulting two-body quasiparticle interactions makes these effective Hamiltonians directly encodable with local qubit operators, opening a concrete path toward their implementation on near-term digital quantum devices through algorithms such as variational quantum eigensolvers and quantum phase estimation.

Whereas the classical quasiparticle BW method achieves high accuracy for all $sd$-shell nuclei studied, the truncated HF BW approach provides quasiparticle effective Hamiltonians that significantly capture correlations in open-shell nuclei. We have tested our results at the level of ground-state energy and fidelity, finding energy errors below 2\% and fidelities exceeding 0.9 in most cases. Therefore, our approach is promising for describing open $sd$-shell nuclei using quantum platforms.
 
This analysis opens the way to a more precise implementation of the quasiparticle approach for nuclear shell model Hamiltonians. Future investigations can test the effective Hamiltonian presented in this work with quantum algorithms such as unitary coupled cluster~\cite{td9s-z7my,Stevenson25} or ADAPT-VQE~\cite{perez2023nuclear,17n2-xh6k} and explore the effect of the truncated BW component in a quantum simulation. In addition, we aim to explore more efficient truncations of the quasiparticle BW method. For instance, different techniques for low energy states, such as gadget Hamiltonians~\cite{PhysRevA.91.012315}, may simplify the eventual encoding of the quasiparticle BW method in a quantum device.

\begin{acknowledgments}
E.C. acknowledges Alba Cervera-Lierta, Caroline Robin and Bruno Juliá-Diaz for interesting discussions and suggestions. He also acknowledges Arnau Rios for the guidance and support throughout this work. This work has been financially supported by the Ministry of Economic Affairs and Digital Transformation of the Spanish Government through the QUANTUM ENIA project call – Quantum Spain project, and by the European Union through the Recovery, Transformation and Resilience Plan – NextGenerationEU within the framework of the Digital Spain 2026 Agenda; and by MCIN/AEI/10.13039/501100011033 from the following grants: PID2023-147112NB-C22, CNS2022-135716 funded by the “European Union NextGenerationEU/PRTR”, and CEX2024-001451-M to the “Unit of Excellence María de Maeztu 2025-2031” award to the Institute of Cosmos Sciences.
\end{acknowledgments}

\bibliography{biblio}

\end{document}